\documentclass{pasj00}

\begin{document}
\SetRunningHead{Kuno et al.}{ }
\Received{2007/06/22}
\Accepted{2007/11/08}

\title{Central Spiral Structure of Molecular Gas in Maffei 2}

\author{Nario \textsc{Kuno},\altaffilmark{1,2} 
Kouichiro \textsc{Nakanishi},\altaffilmark{1,2} \\
Kazuo \textsc{Sorai},\altaffilmark{3} 
and Toshihito \textsc{Shibatsuka},\altaffilmark{4}
}
\altaffiltext{1}{Nobeyama Radio Observatory\thanks{Nobeyama Radio Observatory 
(NRO) is a division of the National Astronomical Observatory of Japan (NAOJ) under the National Institutes of Natural Sciences (NINS).}, Minamimaki-mura, Minamisaku-gun, Nagano 384-1305, Japan}
\email{kuno@nro.nao.ac.jp}
\altaffiltext{2}{The Graduate University for Advanced Studies (SOKENDAI), \\
2-21-1 Osawa, Mitaka, Tokyo 181-0015, Japan}
\altaffiltext{3}{Division of Physics, Graduate School of Science, Hokkaido University, Sapporo 060-0810, Japan}
\altaffiltext{4}{Department of Astronomy, School of Science, The University of Tokyo, \\
 Bunkyo-ku, Tokyo 113-0033, Japan}


\maketitle

\begin{abstract}

Distribution and kinematics of molecular gas in the central region of the barred spiral galaxy Maffei 2 were investigated using a data set of $^{12}$CO(1--0), $^{12}$CO(2--1), CS(2--1) lines and 103 GHz continuum. 
We found that the offset ridges along the kpc-scale bar continue to the central spiral structure embedded in the weak oval structure which is regarded as $x$2 orbit in the bar potential. 
The spiral structure continues toward the center diverging from the oval structure.
The size of these structures is less than $R \sim 100$ pc. 
The mass concentration within $R =$ 35 pc is estimated to be $2 \times 10^{8} M_{\odot}$.
The high mass concentration is consistent with theoretical predictions concerning the creation of such a nuclear spiral structure.
A comparison with the tracers of dense gas and star-forming region suggests that the dense molecular gas traced by CS(2--1) line is formed at the crossing points of $x$1 and $x$2 orbits and the star-forming region appears after $2 \times 10^{5}$ yr which is comparable with the free-fall time of the dense gas traced by the CS line ($\sim 10^{5}$ cm$^{-3}$).

\end{abstract}

\section{Introduction}

To maintain activities of central starbursts or active galactic nuclei in galaxies, fueling of molecular gas toward the center is necessary. 
Bar is one of the possible mechanisms, for the gas fueling. 
Many numerical simulations of gas motion in the bar potential show that bars can accumulate gas toward the central region (e.g., Athanassoula 1992).
Observational evidence of gas accumulation toward the central kilo-parsec region has also been shown. 
For example, Sakamato et al. (1999) and Sheth et al. (2005) showed that barred spiral galaxies have higher central concentration of molecular gas than non-barred spiral galaxies. 
Furthermore, Kuno et al. (2007) showed that the higher central concentration in barred spirals is made by gas accumulation within a region on the order of the bar length and found a correlation between bar strength and the degree of central concentration.

On the other hand, observations of molecular gas with high angular resolution sufficient to resolve the central structure less than 100 pc are still limited (e.g., Schinnerer et al. 2003; Garc\'ia-Burillo et al. 2003).
Such observations are necessary to understand how the gas flows into the nucleus and relates with central activities.
For example, the size of the twin peaks often seen in the central region of barred galaxies is less than several hundred pc (Kenney et al. 1992). Therefore, we need such high angular resolution to investigate the distribution and dynamics of molecular gas in the central region.

Maffei 2 is a barred spiral galaxy that belongs to the IC 342/Maffei group. 
Since Maffei 2 is located behind the Galactic plane, its optical features are not clear. 
However, the bar and spiral structures can be seen clearly in a near infrared image (Jarrett et al. 2003). 
Many CO observations have so far been done (Ishiguro et al. 1989; Kawabe et al. 1991; Israel, Baas 2003; Mason, Wilson 2004; Kuno et al. 2007). 
The CO map obtained with the Nobeyama 45-m telescope shows structures typical of the distribution of molecular gas in barred spiral galaxies; central peak, offset ridges along the leading side of the large-scale bar, condensations of molecular gas at the end of the bar, and spiral arms from the end of the bar. 
They are symmetric against the center.
The motion of molecular gas in the bar is also consistent with that seen in barred spiral galaxies. 
Namely, the large velocity change is seen across the bar implying that the gas moves along the offset ridges toward the center. 
Since Maffei 2 is one of the nearest barred spiral galaxies, we can investigate the distribution and dynamics of molecular gas in the center and the bar in detail. 
Furthermore, since Maffei 2 has a moderate starburst at the center as suggested from IR and radio observations (Rickard, Harvey 1984; Turner, Ho 1994), observations of Maffei 2 with high angular resolution are helpful to understand the mechanism of the gas fueling toward the galactic center by bars and relation with a starburst phenomenon in barred spiral galaxies.

In this paper, we present the results of $^{12}$CO(1--0), $^{12}$CO(2--1), CS(2--1) lines and 103 GHz continuum observations of Maffei 2 with the Nobeyama Millimeter Array (NMA). 
We discuss the central structure of molecular gas and star formation in Maffei 2.
We adopt 2.8 Mpc as the distance of Mafei 2 throughout this paper (Karachentsev et al. 2003).

\section{Observations}

Observations in $^{12}$CO(1--0), $^{12}$CO(2--1), and CS(2--1) were made from Dec. 2002 to Jan. 2004 using NMA. 
103 GHz continuum data was also obtained simultaneously with $^{12}$CO(1--0).
Figure 1 shows the field of views of NMA on the CO map of Maffei2 obtained with the Nobeyama 45-m telescope (Kuno et al. 2007).
The receiver frontends were SIS mixers. 
System temperatures (DSB) were 100 -- 200 K for CS(2--1) , 150 -- 300 K for CO(1--0), and 200 -- 350 K for CO(2--1). 
The UWBC spectro correlator (Okumura et al. 2000) which covers 512 MHz with 256 channels was used as backend (effective frequency resolution: 4 MHz). 
The bandwidth and the frequency resolution correspond to 1563 km s$^{-1}$ and 12.2 km s$^{-1}$ at 98 GHz, 1322 km s$^{-1}$ and 10.3 km s$^{-1}$ at 115 GHz, 666 km s$^{-1}$ and 5.2 km s$^{-1}$ at 230 GHz.
The data of $^{12}$CO(1--0) and CS(2--1) were obtained with antenna configurations AB, C, and D, while only C and D were used for $^{12}$CO(2--1). 
The field center is ($\alpha_{\rm J2000}$, $\delta_{\rm J2000}$) = (\timeform{2h41m54s.6}, +\timeform{59d36'12''.3}) for all observations.
Phase and amplitude calibrations were performed every 20 -- 25 minutes using a nearby quasar NRAO 150 (0359 + 509). 
The quasar was also used for band-pass calibration. 
The absolute flux of NRAO 150 was measured by comparing with URANUS and NEPTUNE. 
The uncertainty of absolute flux is estimated to be about 20 \%. 
We mapped the $^{12}$CO data with uniform weight, while natural weight was used for CS(2--1) and 103 GHz continuum. 
We achieved $1.7'' \times 1.5''$ synthesized beam in $^{12}$CO(1--0), $2.3'' \times 1.4''$ beam in $^{12}$CO(2--1). 
The beam sizes correspond to 23 pc $\times$ 20 pc and 31 pc $\times$ 19 pc at the distance of Maffei 2. 
For CS(2--1) and 103 GHz continuum, the beam sizes are $5.0'' \times 3.6''$ and $4.9'' \times 3.7''$, which correspond to 68 pc $\times$ 49 pc and 67 pc $\times$ 50 pc, respectively.
The observational data are listed in Table 1.

\section{Results}
\subsection{Total Flux}

The missing fluxes of the lines are estimated by comparing single dish observations. Flux within the field of view of NMA ($\sim 60''$) in the $^{12}$CO(1--0) map is 3200 Jy km s$^{-1}$ after correction for primary beam attenuation. 
On the other hand, the total flux within the $56''$ beam measured with a single-dish telescope is 3513 Jy km s$^{-1}$ (Sargent et al. 1985). 
The missing flux is estimated to be about 10 \%. 
Therefore, most of the total flux was recovered in our $^{12}$CO(1--0) map. 
For $^{12}$CO(2--1), the flux within the field of view of NMA ($\sim 30''$) is 6850 Jy km s$^{-1}$. 
The total flux within the $30''$ beam measured by Sarget et al. (1985) is 14880 Jy km s$^{-1}$.
The total flux of NMA is only 46 \% of the single dish measurement.
However, the total flux within the $21''$ and $43''$ beams measured by Israel and Baas (2003) is 4700 Jy km s$^{-1}$ and 8527 Jy km s$^{-1}$, respectively. 
These results seem to be consistent with ours.
When we convolve our data to the $21''$ beam, the peak flux is 3468 Jy km s$^{-1}$ which corresponds to 74 \% of the single dish measurement.
Total flux of CS(2--1) line is 68 Jy km s$^{-1}$.
Total flux density of 3 mm continuum is 38 mJy.
Unfortunately, there are no single dish data of CS(2--1) and 103 GHz continuum.

\subsection{Integrated Intensity Map}
\subsubsection{$^{12}$CO(1--0) and $^{12}$CO(2--1)}

Figure 2 shows the integrated intensity maps of $^{12}$CO(1--0) and $^{12}$CO(2--1). 
Although the $^{12}$CO(1--0) map is consistent with the previous map (Ishiguro et al. 1989; Kawabe et al. 1991), our map shows more detailed structures. 
In the bar region, the offset ridges along the leading side of the bar can be seen. 
The width of the ridge (FWHM) deconvolved with the beam size is $3.7''$ which corresponds to 50 pc at the distance of Maffei2. 
The offset ridges are truncated at about $20''$ from the center on both sides. 
At the edges of the offset ridges, they have bifurcation toward the trailing side. 
It is interesting that the same feature is seen in the $^{12}$CO(1--0) map of IC 342 (Meier et al. 2000). 
The offset ridges connect to strong peaks at about $5''$ ($\sim 70$ pc) north and south of the center. 
They can be regarded as the same feature as the twin peaks found by Kenney et al. (1992) in some barred spiral galaxies. 
The size of both peaks (FWHM) is about $4'' \times 10''$ (54 pc $\times$ 136 pc). 
The center derived from 2 MASS data (\timeform{2h41m55s.0}, +\timeform{59d36'14".7} (J2000): Jarrett et al. 2003) is located between the peaks. 
At the peaks, the offset ridges, especially the northern ridge, show large curvature. 
Since the inclination angle of Maffei 2 is fairly large ($67^{\circ}$: Hurt et al. 1996), some structures overlap along the line of sight. 
Such structures mixed in the peaks can be separated in the velocity domain, as shown in section 3.2. 

The distribution of $^{12}$CO(2--1) emission is consistent with that of $^{12}$CO(1--0) emission. 
The curve of the ridges at the peaks is more apparent in $^{12}$CO(2--1), especially in the southern ridge. 
Although the size of the peaks is about the same as that in $^{12}$CO(1--0), they are closer to the center. 
In both lines, the emissions are weak at the center.

Assuming a conversion factor from CO(1 -- 0) intensity to column density of H$_{2}$ $N($H$_{2})/I_{\rm CO}$ of $1 \times 10^{20}$ cm$^{-2}$ [K km s$^{-1}$]$^{-1}$ (Nakai, Kuno 1995), the molecular gas mass within the field of view ($R = 407$ pc) is estimated to be about $6.8 \times 10^{7} M_{\odot}$. 
This is smaller than the values reported in the previous observations (e.g., Ishiguro et al. 1989) because of the difference in the adopted distance and the conversion factor. 
If we use the same distance and conversion factor, our result is consistent with previous studies. 
The peak flux of $^{12}$CO(1--0) is 47.3 Jy beam$^{-1}$ km s$^{-1}$ which corresponds to the column density of $N({\rm H_{2}}) = 1.7 \times 10^{23}$ cm$^{-2}$ or the surface density of about 2700 $M_{\odot}$ pc$^{-2}$. 
Since the width of the ridges (FWHM) in the map is about 50 pc, the thickness must be smaller than that. 
If we assume that the thickness and inclination angle of the gas disk are 50 pc and 67 deg, respectively, a lower limit of the average volume density within the beam is derived to be $\sim 400$ cm$^{-3}$. 
The peak flux density of $^{12}$CO(1--0) at the twin peaks is about 600 mJy beam$^{-1}$ which corresponds to $T_{\rm b} = 22$ K. 
On the other hand, the peak flux density of $^{12}$CO(2--1) is about 2.8 Jy beam$^{-1}$ which corresponds to $T_{\rm b} = 20$ K. 
Although these values are the average within the beam and a lower limit of the kinetic temperature of the molecular gas, the temperature is higher than that of GMCs in our Galaxy.
We will discuss the physical properties of molecular gas including observations of other lines in a forthcoming paper.

\subsubsection{CS(2 -- 1)}

Figure 3 shows the integrated intensity map of CS(2--1). 
The CS map also show twin peaks. 
The peaks coincide with the CO peaks.
The northern peak is much stronger than the southern side.

\subsubsection{3mm continuum}

Figure 4 shows the 3 mm continuum map. 
There is a peak about $3''$ north of the center with an extended structure along the declination. There is a weak peak about $6''$ south of the center. 
The overall size is $\sim 4.5'' \times 20''$ ($\sim 60 {\rm pc} \times 270 {\rm pc}$). 
The 3 mm continuum is thought to trace star-forming regions, since it is dominated by free-free emission (Condon 1992). 
Actually, these structures coincide with those in other tracers of star-forming activity, such as the 2 cm and 6 cm continua (Turner, Ho 1994), 10.8 $\mu$m (Telesco et al. 1993). 
However, the correlation with H$\alpha$ is not good. 
The H$\alpha$ peak in Ishiguro et al. (1989) is located about $10''$ north of the 103 GHz peak. 
The reason for the bad correlation between free-free emission and H$\alpha$ is thought to be dust extinction in H$\alpha$. 
A similar trend is seen in M82 (Matsushita et al. 2005). 

The total flux density of the 3 mm continuum is 38 mJy. 
The peak flux density within the $4.9'' \times 3.7''$ beam is about $1/3$ of the total flux density. 
Therefore, most of the emission is concentrated in the peak.
We estimate the production rates of Lyman continuum photons, $N_{\rm lyc}$, using the flux according to Condon (1992). 
The result derived from the total flux density is 
\begin{equation}
N_{\rm lyc} / {\rm s}^{-1} = 3.5 \times 10^{52} \times (T_{\rm e} / 10^{4} {\rm K})^{-0.45} \times (d / 2.8 {\rm Mpc})^{2},
\end{equation}
where $T_{\rm e}$ and $d$ are electron temperature and distance, respectively. 
This is about two times larger than that derived from the 6 cm continuum (Turner, Ho 1994) after correcting the difference of the adopted distance. 
As compared with the starburst galaxy M82, the derived $N_{\rm lyc}$ of Maffei 2 is comparable with that of an individual peak in M82 measured with the $43 {\rm pc} \times 36 {\rm pc}$ beam (Matsushita et al. 2005). 
The massive star formation rate ($M > 5M_{\odot}$) derived from the production rate of Lyman continuum photons using the equation (24) in Condon (1992) is $0.1 M_{\odot} {\rm yr}^{-1}$. 
On the other hand, the star formation rate derived from the infrared luminosity of $4.2 \times 10^{9} L_{\odot}$ (Rickard, Harvey 1984) using the equation (26) in Condon (1992) is $0.1 M_{\odot} {\rm yr}^{-1}$, if the distance we adopted is used. 
They show good agreement.

\subsection{Velocity Structure}

\subsubsection{Channel Map}

Figures 5 and 6 are the channel maps of $^{12}$CO(1--0) and $^{12}$CO(2--1). 
They show quite similar features.
Near the systemic velocity of Maffei 2 (-24 km s$^{-1}$: Kuno et al. 2007), two peaks are seen north and south of the center, shown by a cross in the figures. 
For a circular motion, the emission should be observed near the minor axis at systemic velocity. 
The peaks, however, are located near the major axis, indicating large non-circular motion. 
The peaks move toward the north with decreasing radial velocity keeping the separation of the peaks almost constant. 
On the other hand, they move toward the south with increasing radial velocity. 

\subsubsection{Position-Velocity diagrams}

Position-velocity diagrams are helpful to understand the complex velocity structure in Maffei 2. 
Figure 7 shows the position-velocity diagrams of $^{12}$CO(2--1) parallel to the major axis of Maffei 2 (P.A. = $26^{\circ}$: Hurt et al. 1996). 
It is apparent that the central structure is separated into two parallel components in the P -- V diagrams (B-C-D, B'-C'-D').
It implies that there are two components along the line of sight. 
The radial velocities at the center of both components are not systemic velocity of Maffei 2 (-24 km s$^{-1}$). 
There is no emission at systemic velocity at the center. 
The parallel components are clearly separated in the P--V diagram along the minor axis as shown in Figure 8.
Since the parallel structures in the P--V diagrams are symmetric against the 2 MASS center, the center is thought to coincide with the dynamical center. 
For the northern component (upper side) in Figure 7(c), A' -- B' corresponds to the offset ridges of the large-scale bar (A -- B in Figure 7(a) for the southern component). 
If we assume that the spiral arms of Maffie 2 are trailing arm, the eastern side against the major axis of the galaxy is far side and the large-scale bar approaches us closing to the center for the northern side of the galaxy (Figure 1).  
Therefore, if the gas moves along the large-scale bar, the observed blue-shift of the radial velocities from the systemic velocity means inward motion.
The offset ridge connects to the parallel structure at B' (B for the southern component).
The radial velocity along the parallel structure changes from blue-shift to red-shift before the structure crosses the minor axis (B' -- C'), although the point should be on the minor axis for a circular motion. 
The same feature but in the opposite sense can be seen for the southern component. 
Furthermore, the parallel components are connected by a weak oval structure, as indicated in Figure 7b. 
There are bifurcations in the oval structure at C and C'.
The radial velocity increases closing to the center from the points to D and D'.

\section{Discussion}

\subsection{Nuclear spiral structure of molecular gas}

As mentioned in the previous section, the P -- V diagrams along the major axis (Figure 7) shows two components almost overlapping along the line of sight. 
These structures can be regarded as a spiral structure embedded in a weak oval structure. 
We made integrated intensity maps of each component of the parallel structure in Figure 7 separately using the channel map. 
Figure 9 shows the integrated intensity map overlaid on the HST image (F814W) obtained from the HST Archive.

They look like the two-arm spiral structure continues from the offset ridges along the large-scale bar in the CO map obtained with the NRO 45-m telescope. 
The structure can be seen in both $^{12}$CO(1--0) and $^{12}$CO(2--1). 
The spiral structure is symmetric against the center determined from the 2MASS data. 
The northern arm coincides with the dust lane in the HST image very well near the center, while the dust lane along the southern arm is not clear.
This is because the northern arm is on the near side (western side against the major axis of the galaxy) near the center.

It is apparent that the oval structure in Figure 7b is not an expanding ring. The part of the oval structure that overlaps with the northern arm is on the near side near the center, as mentioned above. The radial velocity of the part is red-shift from the systemic velocity of Maffei 2 (around C' in figure 7c).
On the other hand, if the ring is shrinking, the velocity toward the center must be higher than 50 km s$^{-1}$. 
In that case, the oval structure collapses into the nucleus in less than $2 \times 10^{6}$ yr.
This seems to be unlikely. 
The oval structure can be naturally understood as an oval motion of molecular gas.
The oval structure connects the spiral structure.
The schematic figure of the spiral and oval structures is shown in Figure 10. 
Here, we assume inclination and position angle derived for the galactic disk. 
Similar structures with the offset ridges and oval structure are often seen in numerical simulations (e.g., Athanassoula 1992). 
These structures are made from $x$1 and $x$2 orbits in a bar potential. 
Similar structure has been found in NGC 4303 (Schinnerer et al. 2002; Koda, Sofue 2006).
The complex P--V diagrams can be understood with our interpretation.
The offset ridges along the large-scale bar contact with the x2 orbit at B and B' in Figure 7.
The constant gradient of the radial velocity B -- C and B' -- C' in Figure 7 is attributed to the flow of the molecular gas along the spiral arms around the center. 
The spiral arms diverge from the oval structure toward the center.
This corresponds to C -- D and C' -- D' in the P -- V diagrams.
The deviation from the systemic velocity becomes max there, since the gas stream line closes to the line of sight. 

The sizes of the structure shown in Figure 10 were derived from the P -- V diagrams along the major and minor axes and the channel map as follows:
(1) The separation along the major axis between the spiral arms derived from Figure 7(b) ($\sim 70$ pc).
(2) The separation along the minor axis between the spiral arms derived from Figure 8 ($\sim 90$ pc).
(3) The separation along the major axis direction between the positions where radial velocity equals to the systemic velocity. If we assume that most of the gas moves along the spiral arms, the spiral arms are perpendicular to the line of sight at these positions. This can be measured in channel map at the systemic velocity of Maffei 2 ($\sim$ -24 km s$^{-1}$) (Figure 5 and 6). ($\sim 75$ pc).
(4) The separation along the minor axis between the positions where radial velocity equals the systemic velocity derived from Figures 5 and 6. ($\sim 120$ pc).
(5) The length of the oval structure along the major axis derived from Figure 7(b) ($\sim 200$ pc).

\subsection{Formation mechanism of the spiral structure}

Martini et al. (2003a; 2003b) show from HST images that galaxies with strong bars often have a grand-design nuclear dust spiral structure, which appears to be the continuation of the dust lanes along the leading edges of a large-scale bar toward the center. 
On the other hand, observations of such spiral structures of molecular gas are still rare.
The distribution of molecular gas in Maffei 2 can be regarded as such a spiral structure. 
Although Maffei 2 is classified as SAB(rs) in RC3, it seems to be due to the unclear optical image. 
According to Buta and Block (2001), Maffei 2 should be classified as SB(rs). 
Therefore, our result is consistent with Martini(2004)'s finding that grand-design spiral structure is only found in SB(s) or SB(rs). 
Some numerical simulations show that when there is a massive concentration at the center, such as a super massive black-hole, a nuclear spiral structure is formed (Fukuda et al. 1998; Ann, Lee 2004; Maciejewski 2004). 
This is because the mass concentration removes the rigid rotation part from the rotation curve. 
As a result, IILR (inner Inner Lindblad Resonance), where a circum-nuclear ring is formed, disappears or another resonance (Nuclear Lindblad Resonance) appears inside the IILR. 
The mass of the central concentration assumed in numerical simulations to make the spiral structure is 0.1 \% -- 1 \% of the total mass of the host galaxies. 
The observational data of molecular gas has the advantage that it can be used to investigate the kinematics of the central structure.
From the P -- V diagram along the major axis (Figure 7), the rotation velocity is estimated to be about 140 km s$^{-1}$ at $R = 35$ pc. 
The central mass within $R =$ 35 pc is estimated to be $2 \times 10^{8} M_{\odot}$ from the rotation velocity assuming spherical distribution of mass.  
The mass is about 0.5 \% of the total dynamical mass of Maffei 2 (Mason, Wilson 2004). 
Even if the rotation velocity is overestimated due to the non-circular motion, it seems to be more than 0.1 \% of the dynamical mass (Koda, Wada 2002).
The mass fraction of the central mass concentration in Maffei 2 is comparable to the prediction of the simulations. 
The rotation velocity still rising toward the center at $R = 35$ pc.
It is desirable to obtain the inner rotation curve with higher spatial resolution and higher sensitivity observations to know the mass distribution in the inner region.

\subsection{Inflow of Molecular Gas along the Bar}

The molecular gas changes the direction of motion with a large velocity change when it enters the offset ridge along the large-scale bar. 
Then, the gas flows toward the center along the offset ridge. 
Since the radial velocity is almost constant along the straight ridges, as seen in Figure 7, the streaming velocity along the ridge seems to be constant. (The component of the pattern speed of the bar is relatively small as compared with the velocity along the bar.)
The velocity along the ridge is estimated to be 70-120 km s$^{-1}$ assuming the inclination angle and position angle of the galactic disk. 
Here, we assumed a pattern speed of the bar of 53.5 km s$^{-1}$ kpc$^{-1}$ which was adopted from Hurt et al. (1996) and made a correction for the difference of the adopted distance. 
The velocity is comparable with those in NGC 253 (Sorai et al. 2000) and NGC 1530 (Regan et al. 1997). 
Using the velocity and surface density of molecular gas at the radius of 200 pc in the offset ridge, the mass flow of molecular gas along the ridges is derived to be 7--12 $M_{\odot} {\rm yr}^{-1}$. 
This is comparable with the estimation for NGC 1530 by Regan et al. (1997).

If the molecular gas in the flow is accumulated in the center directly, it takes only $< 10^{8}$ yr to concentrate all of the molecular gas in the disk on the oval structure.
However, only a portion of the flowing gas must be accumulated in the central region as estimated by Regan et al. (1997).
They evaluated it to be 20 \% of the flowing gas.

\subsection{Central starburst in barred spiral galaxies}

There are concentrations of molecular gas where the offset ridges connect to the spiral structure (Figures 7 and 9). 
The mass of the condensations within 65 pc $\times$ 94 pc is estimated to be $\sim 7 \times 10^{6} M_{\odot}$ from the $^{12}$CO(1--0) intensity using a conversion factor of $1 \times 10^{20}$ cm$^{-2}$ [K km s$^{-1}$]$^{-1}$ (Nakai, Kuno 1995). 
The peaks correspond to $x$1/$x$2 orbit-crowding region.

We compare the position of the peaks with the peaks in the CS(2--1) and 3 mm continuum maps.
Since the northern peak is much stronger than the southern peak in the CS(2--1) and 103 GHz maps, we discuss the northern peak here. 
The peak of the 3mm continuum which is a tracer of star-forming regions is located downstream of the CO peak. 
The offset between the CO and 3 mm peaks is about 20 pc, while the CS peak coincides with the CO peak.
We can estimate the time scale for the gas to move from the CS peak to the 3 mm peak along the ridge using the streaming velocity measured at the offset ridge, 70 -- 120 km s$^{-1}$, assuming that the peaks are near the point where the spiral structure is perpendicular to the line of sight (cf. Figure 10). 
The time scale is about 1.6 -- 2.8 $\times 10^{5}$ yr. 
On the other hand, free fall time of the dense gas traced by CS ($\sim 10^{5}$ cm$^{-3}$) is estimated to be $1.4 \times 10^{5}$ yr. 
This is comparable to the time for the gas to move from the CS peak to the 3 mm peak. 
From these results, the following scenario can be described:
The molecular gas in $x$1 orbit moves along the offset ridge in the large bar. 
At the $x$1/$x$2 orbit-crowding region, the gas contacts with $x$2 orbit and is compressed by shock and the density rises abruptly there. 
Once the dense gas is formed, it evolves into stars with free-fall time. 
It is interesting to compare the structure with that seen in NGC 6951 (Kohno et al. 1999). 
Kohno et al. (1999) show that the condensations of dense gas traced by HCN in NGC 6951 are located downstream of more diffuse gas traced by CO. 
They showed that the time scale of gravitational instability and flowing time of the gas along the ridges are comparable ($\sim 1 \times 10^{6}$ yr). 
This result supports the idea that the dense gas is made by gravitational instability during the flow along the ridge. 
We speculate that the difference of our results from Kohno et al. (1999) may be caused by the difference in the size of the structure. 
The structure found in NGC 6951 is several times larger than that in Maffei 2. 
The size of the CO peak is more than hundreds pc, while the condensations in Maffei 2 are comparable to the Galactic GMCs. 
Therefore, the molecular clouds should be treated as particles in NGC 6951. 
In that case, since the viscosity of the gas is small, shock may not play an important role in dense gas formation at the $x$1/$x$2 orbit-crowding region, while the gravitational instability does. 
On the other hand, the clump of molecular gas whose size is comparable with a GMC is compressed at the crossing point in Maffei 2. 
As a result, strong shock may occur and dense gas is formed at a burst there. 

\section{Summary}

We made $^{12}$CO(1--0), $^{12}$CO(2--1), CS(2--1) lines and 103 GHz continuum observations of Maffei 2. 
Using the data set, we investigated the distribution and kinematics of molecular gas in the central region.
The results are summarized as follows: \\
(1) We revealed the spiral structure of molecular gas in the central region. The offset ridges of molecular gas along the leading side of the large-scale bar continue to the spiral structure embedded in the weak oval structure which is regarded as a x2 orbit.
The size of these structures is less than $R =$ 100 pc. 
The spiral structure continues toward the center at least until $R =$ 35 pc diverging from the oval structure. \\
(2) The central mass within $R =$ 35 pc is estimated to be $2 \times 10^{8} M_{\odot}$ from the rotation curve, which corresponds to 0.5 \% of the dynamical mass of Maffei 2. The percentage is consistent with the theoretical predictions to make such a spiral structure of molecular gas. \\
(3) The amount of the molecular gas which flows along the offset ridges of the large-scale bar is estimated to be 7--12 $M_{\odot} {\rm yr}^{-1}$. 
This is an upper limit of the mass flow of molecular gas into the nucleus. \\
(4) Our results imply that dense gas is formed at the crossing points of $x$1 and $x$2 orbits. Massive stars are formed from the dense gas and star-forming regions appear after free-fall time of the dense gas ($ 1 \times 10^{5}$ yr). 

\bigskip

We thank the NMA staff for their kind support and encouragement.
This research used the facilities of the Canadian Astronomy Data Centre operated by the National Research Council of Canada with the support of the Canadian Space Agency.


\begin{figure}
\begin{center}
\end{center}
\caption{Left is the $J + H + K$ image of Maffei 2 from the 2MASS Large Galaxy Atlas (Jarrett et al. 2003). Right is the field of views of NMA on the CO map obtained with the NRO 45-m telescope (Kuno et al. 2006). Larger and smaller circles are those of $^{12}$CO(1--0) and $^{12}$CO(2--1), respectively. The arrow in the right image indicates the position angle of the major axis of Maffei 2.}
\label{}
\end{figure}

\begin{figure}
\begin{center}
\end{center}
\caption{(a) Integrated intensity map of $^{12}$CO(1--0) without primary beam correction. Contour interval is 3 Jy beam$^{-1}$ km s$^{-1}$ which corresponds to 3 $\sigma$. The arrow indicates the position angle of the major axis of Maffei 2. (b) Integrated intensity map of $^{12}$CO(2--1) without primary beam correction. Contour interval is 15 Jy beam$^{-1}$ km s$^{-1}$ which corresponds to 1.5 $\sigma$. Cross indicates the center derived from 2MASS data (Jarrett et al. 2003). Circle indicates the field of view of NMA (50 \% level). }
\label{}
\end{figure}

\begin{figure}
\begin{center}
\end{center}
\caption{Integrated intensity map of CS(2--1). Contour interval is 0.5 Jy beam$^{-1}$ km s$^{-1}$ which corresponds to 3 $\sigma$. 
Cross indicates the center derived from 2MASS data. 
Circle indicates the field of view of NMA (50 \% level).}
\label{}
\end{figure}

\begin{figure}
\begin{center}
\end{center}
\caption{3 mm continuum map. Contour interval is 3 mJy beam$^{-1}$ which corresponds to 1.5 $\sigma$. Cross indicates the center derived from 2MASS data. Circle indicates the field of view of NMA (50 \% level).}
\label{}
\end{figure}

\begin{figure}
\begin{center}
\end{center}
\caption{Channel map of $^{12}$CO(1--0). Contour interval is 120 mJy beam$^{-1}$ which corresponds to 3$\sigma$. 
Central velocities ($V_{\rm LSR}$ in km s$^{-1}$) are labeled in upper right. 
Velocity width of each channel is 10.3 km s$^{-1}$. 
Cross indicates the center derived from 2MASS data.}
\label{}
\end{figure}

\begin{figure}
\begin{center}
\end{center}
\caption{Channel map of $^{12}$CO(2--1). Contour interval is 390 mJy beam$^{-1}$ which corresponds to 3$\sigma$. 
Central velocities ($V_{\rm LSR}$ in km s$^{-1}$) are labeled in upper right. 
Velocity width of each channel is 7.8 km s$^{-1}$. 
Cross indicates the center derived from 2MASS data.}
\label{}
\end{figure}

\begin{figure}
\begin{center}
\end{center}
\caption{P -- V diagrams of $^{12}$CO(2--1) along the major axis of Maffei 2 and parallel to the major axis. 
(a) $0.9''$ eastern side of the major axis. 
(b) Along the major axis. 
(c) $0.9''$ western side of the major axis. 
Horizontal line indicates the minor axis. 
Vertical line indicates the systemic velocity of Maffei 2. 
Contour interval is 0.3 Jy beam$^{-1}$.}
\label{}
\end{figure}

\begin{figure}
\begin{center}
\end{center}
\caption{P--V diagram of $^{12}$CO(2--1) along the minor axis of Maffei 2. Contour interval is 0.3 Jy beam$^{-1}$.}
\label{}
\end{figure}

\begin{figure}
\begin{center}
\end{center}
\caption{Integrated intensity map of $^{12}$CO(1--0) of the two components. 
Contour interval is 4 Jy beam$^{-1}$ km s$^{-1}$. 
Color image is the HST image (F814W).}
\label{}
\end{figure}

\begin{figure}
\begin{center}
\end{center}
\caption{(a) Schematic view of the nuclear spiral structure overlapped with the weak oval structure (face-on). 
The sizes are measured observationally (see text). 
(b) The schematic view overlaid on the contour maps of the central structure correcting the inclination of the galaxy.}
\label{}
\end{figure}

\begin{table}
\begin{center}
\caption{Observational data and results}
\begin{tabular}{ccccc}
\hline\hline

 & CS(2--1) & 103 GHz cont. & $^{12}$CO(1--0) & $^{12}$CO(2--1) \\
\hline
Array configuration & AB,C,D & AB,C,D & AB,C,D & C,D \\
Velocity coverage (km s$^{-1}$) & 1563 & --- & 1322 & 666 \\
Velocity resolution (km s$^{-1}$) & 30.5 & --- & 10.3 & 7.8 \\
Synthesized beam ($''$, $^{\circ}$) & $5.0 \times 3.6, -54 $ & $4.9 \times 3.7$, -35 & $1.7 \times 1.5$, 27 & $2.3 \times 1.4$, -34 \\
Synthesized beam (pc) & $68 \times 49$ & $67 \times 50$ & $23 \times 20$ & $31 \times 19$ \\
rms noise of channel map (mJy beam$^{-1}$) & 8 & 2 & 40 & 130 \\
Total flux within field of view (Jy km s$^{-1}$) & 68 & 38\footnotemark[$*$] & 3200 & 6850 \\
\hline
\multicolumn{5}{l}{\hbox to 0pt{\parbox{85mm}{\footnotesize
\par\noindent
\footnotemark[$*$] total flux density (mJy). 
}\hss}}
\end{tabular}
\end{center}
\end{table}

\end{document}